\documentclass[a4paper,aps,pre,twocolumn,superscriptaddress,groupedaddress]{revtex4}  
\usepackage[margin=1in]{geometry}
\usepackage{natbib}
\usepackage{graphicx}  
\usepackage{makecell}
\usepackage{dcolumn}   
\usepackage{bm}        
\usepackage{amssymb,amsmath}   
\usepackage{array,float}
\usepackage{color}
\usepackage{float}
\usepackage{subfigure}
\usepackage{verbatim}
\usepackage{cleveref}
\usepackage{pdfsync}
\usepackage{wrapfig}
\usepackage{comment}
\usepackage[version=4]{mhchem}
\usepackage[export]{adjustbox}

\newcommand{\be}{\begin{equation}} \newcommand{\ee}{\end{equation}}
\newcommand{\ba}{\begin{eqnarray}} \newcommand{\ea}{\end{eqnarray}}

\newcommand{\III}{\text{I/II}}
\newcommand{\I}{\text{I}}
\newcommand{\II}{\text{II}}

\def\bg#1\eg{%
    \begin{gather}%
    #1%
    \end{gather}%
}
\newcommand{\eq}[1]{\mbox{Eq. (\ref{eq:#1})}}

\def\showjon{} 
\ifdefined\showjon
\newcommand{\jon}[1]{\color{blue}  #1  \color{black}}
\else
\newcommand{\jon}[1]{}
\fi

\ifdefined\showpmm
\newcommand{\pmm}[1]{\textcolor{purple}{#1}}
\else
\newcommand{\pmm}[1]{}
\fi

\newcommand{\qd}{\, .}
\newcommand{\qc}{\, ,}

\makeatletter
\newcommand*{\balancecolsandclearpage}{%
  \close@column@grid
  \clearpage
  \twocolumngrid
}
\makeatother

\begin{document}
\title{
Chemical Kinetics and Mass Action
in Coexisting Phases
%

%
}
\affiliation{Max Planck Institute for the Physics of Complex Systems,
N\"{o}thnitzer Strasse~38, 01187 Dresden, Germany}
\affiliation{Center for Systems Biology Dresden,  Pfotenhauerstrasse~108, 01307 Dresden, Germany}
\affiliation{Max Planck Institute of Molecular Cell Biology and Genetics, Pfotenhauerstrasse 108, 01307 Dresden, Germany}

\affiliation{Cluster of Excellence Physics of Life, TU Dresden, 01062 Dresden, Germany}

\affiliation{Faculty of Mathematics, Natural Sciences, and Materials Engineering: Institute of Physics, University of Augsburg, Universit\"atsstr. 1, 86159 Augsburg, Germany}
\affiliation{contributed equally}
\affiliation{Corresponding authors: {julicher@pks.mpg.de} and
{christoph.weber@physik.uni-augsburg.de}}

\author{Jonathan Bauermann}
\affiliation{Max Planck Institute for the Physics of Complex Systems,
N\"{o}thnitzer Strasse~38, 01187 Dresden, Germany}
\affiliation{Center for Systems Biology Dresden,  Pfotenhauerstrasse~108, 01307 Dresden, Germany}
\affiliation{contributed equally}

\author{Sudarshana Laha}
\affiliation{Max Planck Institute for the Physics of Complex Systems,
N\"{o}thnitzer Strasse~38, 01187 Dresden, Germany}
\affiliation{Center for Systems Biology Dresden,  Pfotenhauerstrasse~108, 01307 Dresden, Germany}
\affiliation{contributed equally}

\author{Patrick M. McCall}
\affiliation{Max Planck Institute for the Physics of Complex Systems,
N\"{o}thnitzer Strasse~38, 01187 Dresden, Germany}
\affiliation{Center for Systems Biology Dresden,  Pfotenhauerstrasse~108, 01307 Dresden, Germany}
\affiliation{Max Planck Institute of Molecular Cell Biology and Genetics, Pfotenhauerstrasse 108, 01307 Dresden, Germany}

\author{Frank J\"ulicher}
\affiliation{Max Planck Institute for the Physics of Complex Systems,
N\"{o}thnitzer Strasse~38, 01187 Dresden, Germany}
\affiliation{Center for Systems Biology Dresden,  Pfotenhauerstrasse~108, 01307 Dresden, Germany}
\affiliation{Cluster of Excellence Physics of Life, TU Dresden, 01062 Dresden, Germany}
\affiliation{Corresponding authors: {julicher@pks.mpg.de} and
{christoph.weber@physik.uni-augsburg.de}}
\author{Christoph A. Weber}
\affiliation{Max Planck Institute for the Physics of Complex Systems,
N\"{o}thnitzer Strasse~38, 01187 Dresden, Germany}
\affiliation{Center for Systems Biology Dresden,  Pfotenhauerstrasse~108, 01307 Dresden, Germany}
\affiliation{Faculty of Mathematics, Natural Sciences, and Materials Engineering: Institute of Physics, University of Augsburg, Universit\"atsstr. 1, 86159 Augsburg, Germany}
\affiliation{Corresponding authors: {julicher@pks.mpg.de} and
{christoph.weber@physik.uni-augsburg.de}}


\begin{abstract}
The kinetics of chemical reactions are determined by the law of mass action, which has been successfully applied to homogeneous, dilute mixtures.
At non-dilute conditions, interactions among the components can give rise to coexisting phases, which can significantly alter the kinetics of chemical reactions.
Here, we derive a theory for chemical reactions in coexisting phases at phase equilibrium.
We show that phase equilibrium couples the rates of chemical reactions of components with their
diffusive exchanges between the phases.
Strikingly, the chemical relaxation kinetics can be represented as a flow along the phase equilibrium line in the phase diagram.
A key finding of our theory is that
differences in reaction rates between coexisting phases stem solely from phase-dependent reaction rate coefficients.
Our theory is key to interpret how concentration levels of reactive components in condensed phases control chemical reaction rates in synthetic and biological systems.
\end{abstract}
\maketitle


\section*{Significance statement}

Chemical reactions in biological systems occur in condensed, heterogeneous environments which can for instance originate from phase separation.
Up to now, the theory of chemical kinetics focused on single-phase systems while a full understanding of the interplay between chemical processes and phase separation was lacking.
Here, we present a theory of chemical processes at phase equilibrium.
We show that increasing the local concentration of reactive components may not increase the reaction rate at phase equilibrium -- a behavior that differs from the classical mass-action law for dilute mixtures.
This unconventional chemistry that we uncover in coexisting phases is key to unravel how condensed phases control
biochemical processes in living cells.

\section{Introduction}

The law of mass action sets the foundation for the kinetics of chemical reactions.
It states that the rate of a reaction is proportional to the chemical activities of the reactants involved where the chemical activity depends on concentrations.
For a bimolecular reaction between the reactants $A$ and $B$ in a dilute, homogeneous solution, the reaction rate $r$ is proportional to the concentrations of both reactants, $r\propto n_A n_B$.
This proportionality arises because the encounters among reactants increase with the concentrations of both components~\cite{benson:1960,christov:1980}.
For a reversible bimolecular reaction, $ A+ B  \rightleftharpoons C+D  \qc$
chemical equilibrium
corresponds to the balance of forward and backward rates.
At dilute conditions and chemical equilibrium,
the law of mass action implies that the ratio of products to reactant equilibrium concentrations is constant, $K\equiv (n_C n_D)/(n_A n_B)$, where $K$ is  the equilibrium constant of the reaction and $n_i$ are the concentrations of the reactive components, $i=A,B,C,D$.

The law of mass action also lays the foundation of reaction-diffusion models. Such models have been used to unravel the minimal principles underlying chemical patterns in non-living~\cite{Kepper:1990,horvath:2009} and living systems.
Examples include pattern formation in tissues~\cite{Turing:1952,asai:1999,meinhardt:1982} 
or on artificial and cellular membranes~\cite{Loose_Schwille_Science_2008, Goehring2011, halatek2018rethinking, Lars2019,Ramm_Schwille_2019}.
In models of such systems, chemical reaction rates and diffusive fluxes are typically considered to be independent~\cite{kondo:2010}.

However, the law of mass action needs to be carefully applied when
mutual interactions among the components become important, particularly in non-dilute solutions, where these interactions couple diffusion and chemical reactions.
Such interactions can also give rise to phase separation, whereby the system demixes to form compositionally distinct coexisting phases.
At phase equilibrium, all components in the coexisting phases have equal chemical potentials, and equivalently, equal chemical activities.
Since chemical activities also determine chemical reactions, the condition of phase equilibrium is expected to govern
chemical equilibrium~\cite{seyboldt:2019}
as well as
chemical reaction kinetics in a way that differs from homogeneous solutions.

Recent experimental studies investigated the effects of coexisting phases on reversible and irreversible chemical reactions~\cite{Strulson2012, Drobot2018, spruijt:2019,Testa:2021,arosio:2021,Peeples:2021}.
Increased~\cite{Strulson2012} as well as decreased~\cite{Drobot2018} reaction rates were reported inside condensed phases compared to their coexisting environment.
Interestingly, an up-concentration of reactive components inside the condensed phase led to a decrease in reaction rates in some cases.
It was suggested that this opposite trend is due to highly composition-dependent reaction rate coefficients~\cite{Drobot2018,spruijt:2019}.
To unravel the physiochemical principles underlying these experimental observations requires a theory of mass-action kinetics in phase separated systems.

Here, we derive the kinetic theory for chemical reactions in coexisting phases that are at phase equilibrium.
A key result of our work is that reaction rates are only different in two phases due to different reaction rate coefficients and not because of density differences leading to different frequencies of encounters among components.
This result stems from the condition of phase equilibrium, which also implies a coupling between diffusion and chemical reaction kinetics due to molecular interactions.
Phase equilibrium allows us to represent the relaxation kinetics of reactions as a chemical trajectory along the binodal manifold in the phase diagram.
When chemical systems are maintained out of chemical equilibrium, we show that specific coexisting phases are selected where reactive components are continuously exchanged between the phases at steady state.


\section{Equilibria in systems with chemical reactions and coexisting phases}

To formulate the theory for chemical reactions in coexisting phase, we first introduce the thermodynamic quantities and discuss the relevant equilibria of such systems.
We emphasize that the following theory is general and can be applied to any type of chemical reaction.
Examples are  complex multicomponent reactions that are common in cellular biochemistry,
which are known to tune protein phase behavior~\cite{Aumiller:2016}. In the following, we focus here on the simplistic $A \rightleftharpoons B$ and $ A + B \rightleftharpoons C$ reactions solely because the reaction kinetics can be visualized.
Furthermore, our theory can also be applied to any  interactions among the reacting components.
To illustrate the interplay between phase separation and chemical reactions, we consider systems where only the product can phase-separate.

\subsection{Chemical potential and  activity}
For a solution composed of different types of
chemical species $C_i$ of molecular volumes $\nu_{i}$
($i=0,...,M$),
the chemical potential
is defined as $\mu_i\equiv \partial G/ \partial N_i|_{T,p,N_{j\neq i}}$, where $G(N_0,N_1,...,N_M,p,T)$ is the Gibbs free energy.
Here, $N_i$ denotes the particle number of chemical species $C_i$, where the index $i=0$ corresponds to the solvent component and $i=1,...,M$ indicate the solute components.
The pressure is denoted by $p$ and $T$ is the temperature.
Introducing the
concentrations $n_i=N_i/V$ with $V=\partial G/ \partial p|_{T,N_{i}}$ denoting the system volume,  we express the chemical potential in terms of the chemical activity
$a_i$~\cite{alberty:2003,burgot:2017,Atkins:2009,Laeuger:2009}:
\be
\mu_i(\{n_k\},p,T) = \mu^0_i(p,T) + k_B T \log(a_i(\{n_k\},p,T)) \qc \label{eq:chem_pot}
\ee
where $\mu^0_i$ is the reference chemical potential for $a_i=1$, and $\{n_k\}$ denotes  $n_0, n_1, ... , n_M$.
The chemical activity, \be
\label{eq:activity}
a_i(\{n_k\},p,T)=  \gamma_i(\{n_k\},p,T) \, n_i \qc
\ee
can be expressed in terms of the activity coefficients $\gamma_i$.
If all solute components are sufficiently dilute with respect to the solvent component, all activity coefficients $\gamma_i$ are positive constants.
In contrast, in non-dilute solutions, interactions among components imply that the activity coefficients $\gamma_i$ depend on the concentrations of all components.
Capturing interactions between components $i$ and $j$ by a mean field energy density $\chi_{ij} N_i N_j/V^2$,
where $\chi_{ij}$ is an interaction parameter, leads to an exponential dependence of the activity coefficient $\gamma_i$ on all the concentrations $n_j$,
\be \label{eq:gamma_specific}
\gamma_i = \nu_i \exp\left( \frac{\sum^{M}_{j=0} \chi_{ij}n_j - \nu_i S}{k_BT} \right)\qc
\ee
with $S=k_B T \sum^{M}_{j=0} n_j + \sum^{M}_{j,k=0}\frac{\chi_{jk}}{2}n_j n_k$ . Note that in the dilute limit the solvent volume fraction $n_0 \nu_0 \simeq 1$, and all interactions involving solute components can be neglected, leading to a constant activity coefficient.
The general form Eq.~\eqref{eq:gamma_specific} of the activity coefficients is consistent with Flory-Huggins model~\cite{flory:1942,Huggins1942}.
We will consider the activity coefficient shown in Eq.~\eqref{eq:gamma_specific}
to study how phase coexistence affects the kinetics of specific chemical reactions.
Note that the corresponding results do not qualitatively depend on the specific form of activity coefficients related to a mean field approximation.


\begin{figure*}[t]
    \centering
    \includegraphics[width=0.99\textwidth]{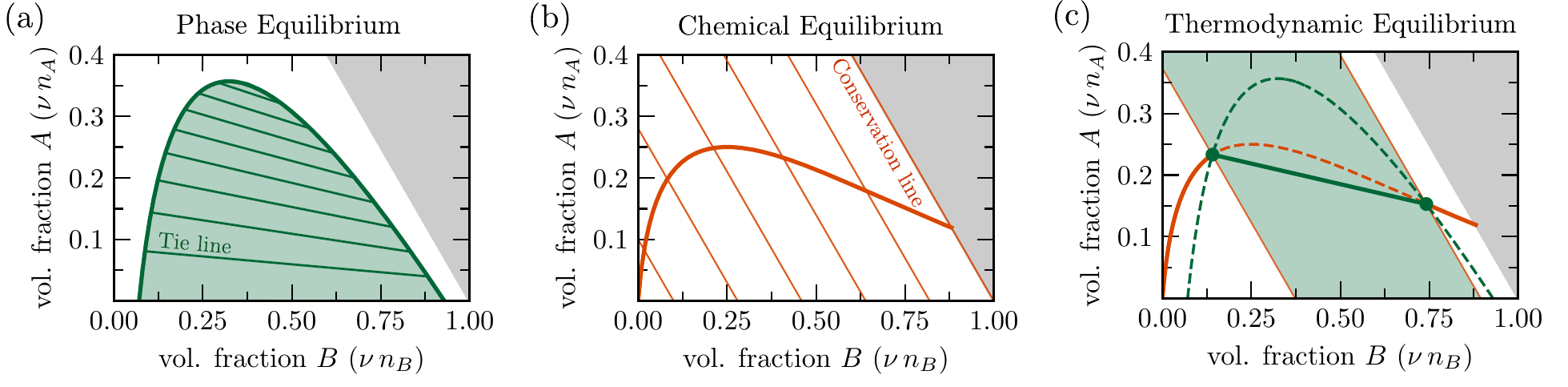}
    \caption{\textbf{Equilibrium phase diagrams.}
    To illustrate different equilibria, we consider a ternary, incompressible mixture composed of the components $A$, $B$ and a non-reactive solvent, with identical molecular volumes $\nu$.
    \textbf{(a)} Phase equilibrium.
    Along the binodal line (thick green line), the phase equilibrium condition~\eqref{eq:phase_eq} is fulfilled.
    Each coexisting pair of equilibrium volume fractions along the binodal line are connected by a tie line (thin green line).
    \textbf{(b)} Chemical equilibrium.
    Along the mono-nodal (thick orange line),
    the chemical equilibrium condition~\eqref{eq:chem_eq} is satisfied.
    Chemical equilibrium for the reaction, $A \rightleftharpoons B$, corresponds to the intersection of the mono-nodal with a conservation line
    $\psi_1=\nu(n_A+n_B)$ (thin orange line).
    \textbf{(c)} Thermodynamic equilibrium.
    In the case of compatible phase and chemical equilibria, both equilibrium conditions (Eqs.~\eqref{eq:phase_eq} and \eqref{eq:chem_eq}) hold simultaneously.
    In this case, a single tie line is selected (thick green line), even for a broad range of conservation lines (green area). For incompatible equilibria, the system is homogeneous and compositions are determined by chemical equilibrium (thick orange line).
    }
    \label{fig:equil}
\end{figure*}

\subsection{Phase equilibrium}

Due to interactions among the components, non-dilute solutions can separate into distinct coexisting phases of different compositions. Though we consider the coexistence of two phases I and II for simplicity, we stress that the theory developed below also applies to coexistence of more than two phases. At phase equilibrium, pressures and temperatures are equal in each phase,  $p^\I=p^\II$ and $T^\I=T^\II$. Also the chemical potentials of each component $i$ balance,
\be
\mu_i^\I = \mu_i^\II \qd \label{eq:phase_eq}
\ee
These conditions are fulfilled by the equilibrium composition of the two coexisting phases with
concentrations $n_i^\III=N_i^\III/V^\III$.
Each pair of equilibrium concentrations, $\{n_k^\I\}$ and $\{n_k^\II\}$, are connected by a tie line. The collection of all such points makes up the binodal manifold in the phase diagram. For instance,  Fig.~(\ref{fig:equil}a) depicts the phase diagram of an example of a ternary mixture.
Equation~\eqref{eq:phase_eq} can be equivalently expressed as an equality of chemical activities in both phases,
\be
a_i^\I=a_i^\II  \qd
\label{eq:phase_eq_activities}
\ee

Molecular species partition unequally in the two phases, which is described by the partition coefficient of species $i$, defined as
\be
P_i \equiv \frac{n_i^\I}{n_i^\II}\qd
\ee
At phase equilibrium \eqref{eq:phase_eq},
the partition coefficients can be expressed in terms of the activity coefficients in both phases, $\gamma_i^\III$, by using Eqs.~\eqref{eq:chem_pot} and \eqref{eq:activity} as
\be
\label{eq:partitioning}
P_i = \frac{\gamma_i^\II}{\gamma_i^\I} \qd
\ee
This expression reveals that partitioning is governed by the composition dependence of activity coefficients $\gamma_i^\III$ in phase separating systems.
If all solutes are dilute with respect to the solvent, solutes partition equally with $P_i=1$ since there is no phase coexistence, i.e., $\gamma_i^\I=\gamma_i^\II$.
Note that the case $P_i=1$ does not necessarily correspond to dilute solutes.

\begin{figure*}[t]
    \centering
    \includegraphics[width=0.99\textwidth]{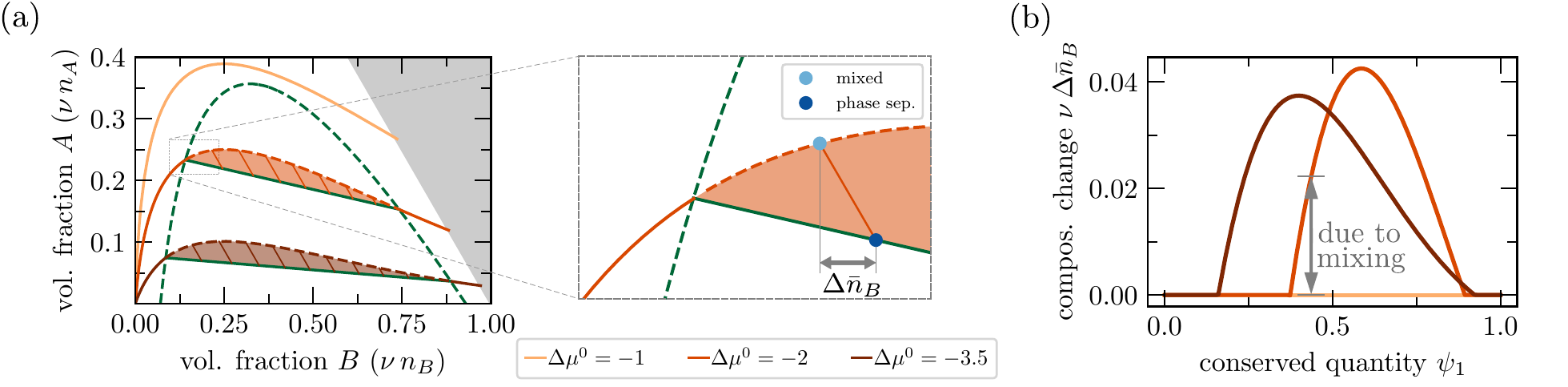}
    \caption{
     \textbf{
     Phase separation controls chemical equilibrium.} \textbf{(a)}
     Varying the difference of reference chemical potential between components $B$ and $A$, $\Delta \mu^0$, leads to different chemical equilibrium lines (orange colored mono-nodals) in the phase diagram.
     The orange lines in the shaded domains (see zoom in) characterize the difference between the phase-separated system at thermodynamic equilibrium and the mixed system (gray arrow), indicating that phase separation can control chemical equilibrium.
     \textbf{(b)} This difference varies with the value of the conserved quantity $\psi_1$, which is depicted by the compositional change due to mixing of the $B$-component, $\nu\Delta\bar{n}_{B}$.
     }
    \label{fig:equil2}
\end{figure*}

\subsection{Chemical equilibrium}

We consider chemical reactions $\alpha=1,...,R$,
\be
\sum_{i=0}^M \sigma_{i\alpha}^+ C_i \rightleftharpoons \sum_{i=0}^M \sigma_{i\alpha}^- C_i \qc \label{eq:chem_scheme}
\ee
of chemical species $C_i$, where $\sigma_{i\alpha}^\pm$ are stoichiometric matrices. In~\eq{chem_scheme}, reactants are on the left side while reaction products are on the right side.
Given $(M+1)$ chemical species undergoing $R$ linearly independent reactions, there exist $(M-R+1)$ conserved quantities $\psi_i$, where $i=0,...,M-R$.
The reaction Gibbs free energies corresponding to reaction $\alpha$ are
\be
\Delta \mu_\alpha \equiv \sum_{i=0}^M \sigma_{i\alpha} \mu_i \qc \label{eq:reac_gibbs}
\ee
where we abbreviate
$\sigma_{i\alpha}=\sigma_{i\alpha}^--\sigma_{i\alpha}^+$.
The condition for chemical equilibrium reads
\be \Delta \mu_\alpha=0\qd \label{eq:chem_eq} \ee
At chemical equilibrium, the concentrations reach equilibrium values
that can be used to define
the
equilibrium reaction coefficients as
\be
K_\alpha \equiv \prod^{M}_{i=0} \left(n_i \right)^{\sigma_{i\alpha}} \qc \label{eq:mass_act}
\ee
which are distinct from the equilibrium reaction constants that include the chemical activities instead of concentrations~\cite{lewis:1907}.

At chemical equilibrium \eq{chem_eq}, the equilibrium reaction coefficients $K_\alpha$ can be expressed in terms of the stoichiometric coefficients, activity coefficients and reference chemical potentials as
\be
K_\alpha =
\prod^{M}_{i=0} \gamma_{i}^{-\sigma_{i\alpha}}\exp\left( -\frac{ \sigma_{i\alpha}\mu^0_i }{k_B T} \right)  \qd
\ee
The equilibrium reaction coefficients thus describe relationships between concentrations at chemical equilibrium.
These  coefficients depend on composition
via the activity coefficients $\gamma_i$~\cite{Atkins:2009,Laeuger:2009}.
For dilute solutions, $K_\alpha$ is composition-independent
and are thus often referred to as equilibrium reaction constants.

In general, for given conserved quantities $\psi_i$,
there is a unique set of concentrations $n_i$ that satisfy
Eq.~\eqref{eq:mass_act} and therefore correspond to chemical equilibrium. Figure~(\ref{fig:equil}a) shows the example of a ternary mixture, where the conserved quantity $\psi_1$ (see figure caption for definition) is constant along the thin straight lines (orange). The concentrations at chemical equilibrium lie on the intersection of a line with constant $\psi_1$ (thin orange line) with the mono-nodal line described by \eq{chem_eq} (thick orange line).

\subsection{Thermodynamic equilibrium}

If a system is at thermodynamic equilibrium and two phases coexist, both chemical reactions and phases are equilibrated.
In this case \eq{chem_eq} and \eq{phase_eq} are obeyed simultaneously.
As a consequence,
thermodynamic equilibrium imposes a relation between equilibrium reaction coefficients and partition coefficients.
In fact, equilibrium reaction coefficients, $K_\alpha^\I$ and $K_\alpha^\II$, differ in the two coexisting phases I and II.
At thermodynamic equilibrium, their ratio obeys
\be
\label{eq:thermod_eq}
\frac{K_\alpha^\I}{K_\alpha^\II}=\prod^{M}_{i=0} \left(P_i\right)^{\sigma_{i\alpha}} \qd
\ee
This equation is a key result of this work since it connects chemical reactions and phase separation at thermodynamic equilibrium.
Equation~\eqref{eq:thermod_eq} can select a subset of coexisting concentrations on the binodal manifold if both equilibria are compatible. The case of compatible equilibria is illustrated in the example of a ternary mixture shown in Fig.~(\ref{fig:equil}c), where a unique pair of concentrations coexist at thermodynamic equilibrium for a large range of conserved quantities (green circles).
Chemical and phase equilibria can also be incompatible. In this case, thermodynamic equilibrium corresponds to a homogeneous state that satisfies only chemical equilibrium Eq.~\eqref{eq:chem_eq}; top orange solid lines in Fig.~(\ref{fig:equil2}a).

An important implication of our theory is that phase coexistence leads to different equilibrium states compared to the corresponding mixed system.
To illustrate this effect,
we compare the average composition $\bar{n}_i$ of a system at phase coexistence to the same system that is mixed but at chemical equilibrium.
Mixing can be realized by stirring for example.
In the phase diagram, this comparison amounts to the deviation between the line of chemical equilibrium (dashed orange line in Fig.~(\ref{fig:equil2}a))
and the tie line (solid green line in Fig.~(\ref{fig:equil2}a)), which is depicted by the orange domains in Fig.~(\ref{fig:equil2}a).

Since the mixed case is only partially equilibrated,
its composition
is different to the composition at thermodynamic equilibrium.
This difference varies with the value of the conserved quantity $\psi_1$ (Fig.~(\ref{fig:equil2}b)).
This dependence on $\psi_1$ solely stems from mixing since
for the considered ternary mixture with one chemical reaction at thermodynamic equilibrium,
changing $\psi_1$ only affects the phase volumes and not composition in each phase.
The difference between the homogeneous,
partially equilibrated state and the phase-separated, thermodynamic state
reflects the influence of phase coexistence on chemical reactions.

\section{Chemical reaction kinetics in coexisting phases at phase equilibrium}

In this section,
we study the kinetics of chemical reactions for systems composed of two homogeneous coexisting phases that are maintained at phase equilibrium but are not at chemical equilibrium.
This condition of a partial equilibrium holds when chemical reactions are slow compared to phase separation and corresponds to the case of a reaction-limited chemical kinetics~\cite{Bar_Even:2011,milo:2015}.

\subsection{Kinetics of concentrations and phase volumes}

In each phase, the kinetics of the respective concentration of component $i$, $n^\III_{i}$ for $i=0,...,M$ is governed by (see Appendix~\ref{app:volume_kinetics})
\begin{subequations}\label{eq:full_kinetics}
\be
\frac{d}{dt}n_i^\III = r_i^\III - j_i^\III - \frac{n_i^\III}{V^\III} \frac{d}{dt}V^\III\qc
\label{eq:dyn_dens}
\ee
where $r_i^\III$ are the reaction rates corresponding to the phase volume $V^\III$ and $j_i^\III$ are the diffusive exchange rates between phases. These rates maintain phase equilibrium at all times. Note that in this work, rates have the units of concentration per time.
The last term of \eq{dyn_dens} accounts for changes in concentrations due to the changes of the respective phase volumes $V^\III$.
The kinetics of these phase volumes follow (see Appendix~\ref{app:volume_kinetics}),
\be
\label{eq:dyn_phasevol}
\begin{split}
\frac{1}{V^\III} \frac{d}{dt}V^\III &= \sum^{M}_{i=0} \nu_i^\III(r_i^\III- j_i^\III) \\& + \sum^{M}_{i=0} n_i^\III\frac{d}{dt}\nu^\III_i \qc
\end{split}
\ee
\end{subequations}
where $\nu_i^\III$ denote the phase-dependent molecular volumes.
If the molecular volumes $\nu_i=\nu_i(p,T)$ are only functions of pressure $p$ and temperature $T$ and are not dependent on composition, then they are equal in both phases at isobaric and isothermal conditions and therefore $d\nu^\III_{i}/dt=0$.
For volume conserving reactions with $\sum^{M}_{i=0} \sigma_{i\alpha} \nu_i =0$, it follows that $\sum^{M}_{i=0} \nu_i r^\III_{i} =0 $ for each reaction $\alpha$.

\subsection{Diffusive exchange rates between phases}
\label{sec:Diff_ex_rates}

To maintain phase equilibrium while chemical reactions occur, components need to be exchanged between the phases.

This exchange conserves the total number of components in the system, which implies for the diffusive exchange rates $j_i^\III$:
\be V^\I j_i^\I = - V^\II j_i^\II \label{eq:con_j} \qd \ee
As a result, for systems with composition independent molecular volumes and volume conserving reactions, the total system volume $V=V^\I+ V^\II$ is constant in time (see Eq.~(\ref{eq:dyn_phasevol}).

The condition of phase equilibrium (Eq.~\eqref{eq:phase_eq_activities}) at all times during the reaction kinetics implies that

\be
\frac{d}{dt}(\gamma_i^\I n_{i}^\I)=\frac{d}{dt}(\gamma_i^\II n_{i}^\II)
\label{eq:maint_phase} \qd
\ee
Using  Eq.~(\ref{eq:full_kinetics}) in Eq.~(\ref{eq:maint_phase}) together with Eq.~(\ref{eq:con_j}), gives a set of $2(M+1)$ equations that are linear in the diffusive exchange rates $j_i^\III$. Therefore, the diffusive exchange rates can be written in closed-form expressions $j_i^\III(\{r_k^\I,r_k^\II\}, V^\I, V^\II)$,
which depend only on the chemical rates $r_k^\III$ and the phase volumes $V^\III$.

\subsection{Reaction rates at phase equilibrium}

The chemical reaction rate of each component, in \eq{dyn_dens}
can be written in terms of the net chemical reaction rate $r_\alpha^\III$ of the reaction $\alpha$ in each of the phases as
\ba
\label{eq:chemcial_rate}
r_i^\III&=& \sum^{R}_{\alpha=1}  \, \sigma_{i\alpha} r_\alpha^\III \ea
This net reaction rate can be split into the forward $(+)$ and the backward $(-)$ reaction rates as $r_\alpha=(r_\alpha^+ - r_\alpha^-)$.

The condition for thermodynamic equilibrium implies that the relationships, $r_\alpha^+=r_\alpha^-$ and  $j_i^\III=0$
hold simultaneously in all phases.
The forward and backward rates in both phases obey detailed balance of the rates,
\be
\label{eq:det_bal_rate}
\frac{r_\alpha^+}{r_\alpha^-}=\exp \left(-\frac{\Delta \mu_\alpha}{k_BT} \right) \qc
\ee
with the reaction free energy $\Delta \mu_\alpha$ given by Eq.~\eqref{eq:reac_gibbs}. Equation~\eqref{eq:det_bal_rate} is fulfilled by choosing the forward and the backward rates as,
\be
\label{eq:chemical_rate_forward_backward}
r_\alpha^\pm = k_\alpha(\{ n_k \}, p,T) \exp \left( \frac{\sum^{M}_{i=0} \sigma_{i\alpha}^\pm \mu_i}{k_BT} \right) \qd
\ee

Here, $k_\alpha (\{n_k \}, p,T)$ denotes a reaction rate coefficient which depends on temperature, pressure and composition $\{n_k \}$. Note that
thermodynamics does not determine the value of the reaction rate coefficient.
Rather, it only constrains the coefficient to be positive and thereby guarantees that the entropy of the system increases.

Using Eq.~\eqref{eq:chemical_rate_forward_backward}, the chemical reaction rate of component $i$ (Eq.~\eqref{eq:chemcial_rate}) can be written as,
\be
\label{eq:prod_rate_with_H}
r_i^\III =  \sum^{R}_{\alpha=1} k_\alpha^\III \sigma_{i\alpha}  H_{\alpha} \qc
\ee
where we introduce the chemical reaction force,
\be
\label{eq:reaction_force_a}
H_{\alpha} =  \exp\bigg({\frac{\mu_\alpha^+}{k_BT}}\bigg) -
 \exp\bigg({\frac{\mu_\alpha^-}{k_BT}}\bigg) \qd
\ee
Here, we have also introduced the forward and backward chemical reaction free energies $\mu_\alpha^\pm$ via $\Delta \mu_\alpha = \mu_\alpha^--\mu_\alpha^+$.

\subsection{Chemical reactions relaxing to chemical equilibrium}

For systems that can relax to thermodynamic equilibrium,
the forward and backward reaction free energies are directly determined by the chemical potentials,
\be\label{eq:mu_pm_TEQ}
\mu_\alpha^\pm = \sum_{i=0}^M \sigma_{i\alpha}^\pm \mu_i
\qc
\ee
and the chemical reaction force can thus be expressed in terms of the  chemical activities as,
\be
\label{eq:reaction_force_b}
H_{\alpha} =  \prod^{M}_{m=0} \left(\text{e}^{\frac{\mu_m^0}{k_BT}}a_m\right)^{\sigma_{m\alpha}^+} -
\prod^{M}_{m=0} \left( \text{e}^{\frac{\mu_m^0}{k_BT}}a_m\right)^{\sigma_{m\alpha}^-} \qd
\ee
This form of the chemical reaction force is specific to systems that can relax to thermodynamic equilibrium and implies various properties for chemical reactions at phase equilibrium.

\subsection{Properties of chemical reactions at phase equilibrium}

First, at phase equilibrium, the chemical activities $a_i=\gamma_i n_i$ are equal in both phases. For chemical reactions that can relax to thermodynamic equilibrium, equal chemical activities between the phases imply that the chemical reaction forces $H_{\alpha}$ (Eq.~\eqref{eq:reaction_force_a}) are equal in both phases as well. Note that the reaction forces are equal despite the composition difference between the phases.
This key result emerges because chemical activities (or equivalently chemical potentials) govern both chemical kinetics of the components in the phases and their diffusion between the phases.
Equal reaction forces $H_{\alpha}$ between phases imply that the component reaction rate $r_i^\III$ shown in Eq.~\eqref{eq:prod_rate_with_H} is different between the phases only due to the composition dependent reaction rate coefficients $k_{\alpha}^\III$.

Second, due to phase equilibrium, the rate of change of the concentration of a reactive molecule in one of the phases, ${d}n_i^\III/{dt}$, is not equal to the chemical reaction rate $r_i^\III$ of the component.
The reason is that, in addition, the exchange of reactive components between the phases $j_i^\III$ and changes in phase volumes ${d}V^\III/{dt}$ contribute to concentration changes in each phase; see Eq.~\eqref{eq:dyn_dens}.
Both contributions are key since they maintain phase equilibrium during the chemical kinetics, i.e., the concentrations $n_i^\III$ remain on the binodal manifold which is defined by the condition for phase equilibrium (Eq.~\eqref{eq:phase_eq}). Thus, determination of reaction rates in each phase requires the knowledge of both the diffusive exchange rates between the phases and how the phase volume changes with time.

Third,
the chemical kinetics at phase equilibrium differs from the kinetics of the corresponding mixed system.
We already discussed in Fig.~(\ref{fig:equil2}a-b) that the
the thermodynamic state is distinct to the corresponding well-mixed system.
In contrast to such well-mixed systems where the chemical kinetics is governed by the composition of the mixture,
the chemical kinetics at phase equilibrium
is determined by the chemical activities (or chemical potentials) along the binodal manifold together with the phase-dependent reaction rate coefficients.
This difference can be illustrated when for example considering the kinetics of the average concentrations, $\bar{n}_{i}=(V^\I\,n^\I_{i}+V^\II\,n^\II_{i})/{V}$.  Using Eqs.~\eqref{eq:full_kinetics}, the corresponding kinetics is given by,
\ba
\label{eq:phi_bar_to_EQ}
\frac{d \bar{n}_i}{dt} =
\sum^{R}_{\alpha=1}  \left(\frac{V^\I}{V} k_\alpha^\I+\frac{V^\II}{V} k_\alpha^\II  \right) \sigma_{i\alpha} H_{\alpha}
\qc
\ea
for volume conserving reactions.
We find that in systems with coexisting phases I and II, the time evolution of the average composition is determined by the kinetics of the phase volumes $V^\I$, the phase-dependent reaction rate coefficients $k_\alpha^\III$ and a phase-independent reaction force $H_{\alpha}$.


\begin{figure*}[t]
  \centering
    \includegraphics[width=0.99\textwidth]{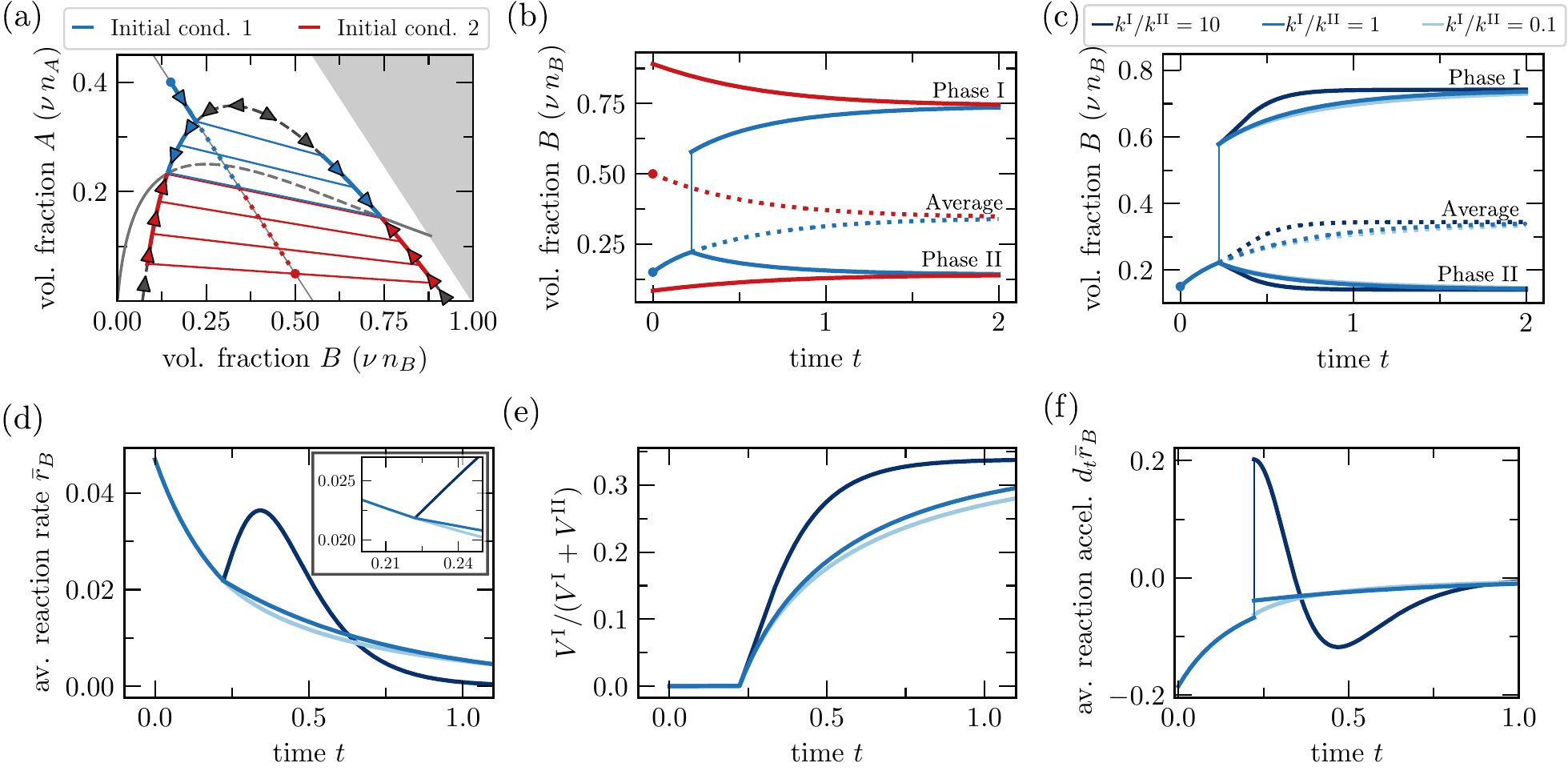}
    \caption{\textbf{Kinetics of a unimolecular reaction relaxing to thermodynamic equilibrium.} We consider the kinetics of a ternary, incompressible mixture and the chemical reaction~\eqref{eq:chem_scheme_unimolecule}.
    \textbf{(a)} Different choices of initial conditions $1$ (blue) and $2$ (red)
    follow different flows fields. However, both initial conditions
    lead to same thermodynamic equilibrium because they lie on the same conserved line $\psi_{1}=0.55$; see also
    \textbf{(b)} showing the chemical trajectories for the product $B$, where solid and dotted lines correspond to volume fraction in phase $\III$ and averages, respectively.
    \textbf{(c)} Difference in reaction rate coefficients between two phases affects reaction rates (in both phases and average) but not the thermodynamic equilibrium state. The average volume fraction of product $B$ changes continuously at the onset of phase separation for initial condition 1.  The average reaction rate of product $B$, $\bar{r}_{B}$ \textbf{(d)}, and the phase fraction, $V^\I/V$ \textbf{(e)}, has a kink at the onset of phase separation, implying that  \textbf{(f)} the average reaction acceleration of product $B$, $d\bar{r}_{B}/dt$, jumps at the onset of phase separation.
    Please note that the kink and the jump require that phase equilibrium is established quasi instantaneously on the time-scales of
    chemical reactions.
    }
    \label{fig:kinetics}
\end{figure*}

\subsection{Unimolecular chemical reactions in coexisting phases}
\label{sec:first_order}

In this section, we discuss an example of a ternary mixture with chemically reactive components $A,B$ and a non-reactive solvent $S$. For simplicity, we assume identical molecular volume $\nu$ for all components.
We consider a single reaction whereby solute $A$ can spontaneously convert to product $B$ and vice versa without the participation of any additional components.
We refer to this reaction as unimolecular chemical reaction,
\be
A  \rightleftharpoons B  \qd \label{eq:chem_scheme_unimolecule}
\ee
Note that for systems that chemically react via unimolecular reactions and that
can phase-separate, component reaction rates Eq.~\eqref{eq:prod_rate_with_H} are generally non-linear in the solute concentrations.
For such unimolecular reaction in a
ternary mixture, we can define two conserved quantities, $\psi_1=\nu(n_A+\nu_{B})$ and $\psi_{0}=\nu n_{S}$.
We numerically solve the governing kinetic equations of the unimolecular chemical reaction Eq.~\eqref{eq:chem_scheme_unimolecule} at phase equilibrium; for details see Appendix~\ref{app:example_ternary}.

\begin{figure*}[t]
  \centering
    \includegraphics[width=0.99\textwidth]{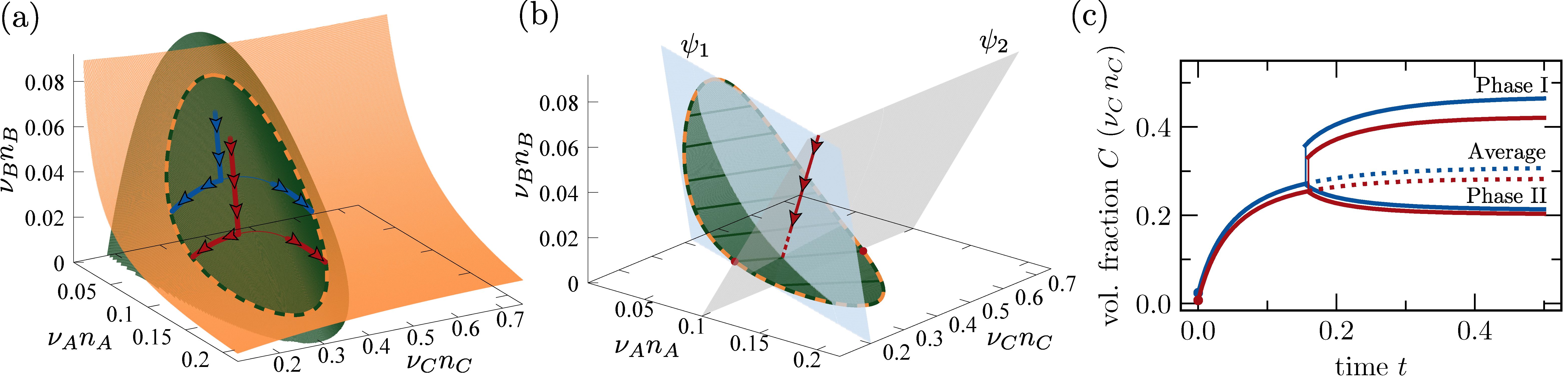}
    \caption{\textbf{Kinetics of a bimolecular reaction relaxing to thermodynamic equilibrium.} We illustrate the kinetics of a quaternary, incompressible mixture
    and the chemical reaction~ Eq.\eqref{eq:chem_scheme_bimolecule}. \textbf{(a)} The phase equilibrium and chemical equilibrium are represented by two surfaces (green and orange, respectively). The closed line given by the intersection of these two surfaces depict the thermodynamic equilibria states (dashed green line). The solid blue and red lines correspond to trajectories of two systems with different values of conserved quantities (see (b)). We have chosen $\psi_1=0.42$ with  $\psi_2=0.05$ for the blue trajectory and $\psi_2=0.1$ for the red trajectory.
    \textbf{(b)} The two planes of conserved quantities defined by $\psi_{1}$ (light blue plane) and  $\psi_{2}$ (gray plane) intersect, to uniquely define a line (solid red as in (a)) along which the trajectory of the average volume fractions progresses during kinetics.
    At the intersection of this unique line with the manifold of thermodynamic equilibrium, the compositions in each phase (red dots) are also uniquely selected. \textbf{(c)} Chemical trajectories for the product $C$, where solid and dotted lines correspond to volume fractions in phase $\III$ and averages, respectively for two systems (red and blue). The volume fractions in each phase and average therefore evolve to different thermodynamic equilibrium states.}
    \label{fig:kinetics_quaternary}
\end{figure*}

The kinetics of a unimolecular chemical reaction at phase equilibrium can be illustrated as a chemical trajectory in a simple phase diagram spanned by two reactive components $A$ and $B$ (Fig.~(\ref{fig:kinetics}a)).
For an initial composition within the binodal (red dot in Fig.~(\ref{fig:kinetics}a)), the concentrations in each phase follow a flow field along the binodal lines (solid red lines in Fig.~(\ref{fig:kinetics}a)). The corresponding average composition moves along the conserved quantity $\psi_1$ while crossing different tie lines (dotted red line in Fig.~(\ref{fig:kinetics}a)).
Changes in tie line as the chemical reaction proceeds imply corresponding compositional changes in the coexisting phases.
For an initial composition outside the binodal (blue dot in Fig.~(\ref{fig:kinetics}a)), the initially well-mixed system
moves along $\psi_1$ and phase separates into coexisting phases when the composition hits the binodal line (solid blue lines in Fig.~(\ref{fig:kinetics}a)).
The onset of phase separation leads to a discontinuity of the volume fractions, which otherwise evolve smoothly in time (Fig.~(\ref{fig:kinetics}b)).
Then, similar to the previous initial condition, the phase composition follows the flow along the binodal lines. Since both cases are identical except for their initial conditions, both relax to the same thermodynamic equilibrium state.

Varying the reaction rate coefficients
$k_\alpha^\III$ in the phases can strongly alter the chemical kinetics  (Fig.~(\ref{fig:kinetics}c)).
When the reaction rate coefficient is increased
in the $B$ product-rich phase ($k_\alpha^\I/k_\alpha^\II=10$),  the product $B$ relaxes more quickly towards thermodynamic equilibrium.
The same holds true for the average concentration of product $B$.
Interestingly, at the onset of phase separation, the average reaction rate $\bar{r}_{i}=(V^\I\,r^\I_{i}+V^\II\,r^\II_{i})/{V}$ (rhs.\ of Eq.~\eqref{eq:phi_bar_to_EQ}) is continuous but can kink for reaction rate coefficients that are unequal between the phases ($k_\alpha^\I \not= k_\alpha^\II$); see Fig.~(\ref{fig:kinetics}d) and inset for average reaction rate of product $B$, $\bar{r}_{B}$.
Average reaction rates $\bar{r}_{B}$ can even initially increase before relaxing to thermodynamic equilibrium ($\bar{r}_{B}=0$).
This increase is a result of an initial very fast growth of phase I, which increases due to the fast formation of product $B$ in phase I (Fig.~(\ref{fig:kinetics}e)).

The kink of the average reaction rate $\bar{r}_{B}$ at the onset of phase separation implies a jump of the acceleration of the chemical reaction, $d \bar{r}_B/ dt$ (Fig.~(\ref{fig:kinetics}f)). In other words, as coexisting phases form, there is a drastic change in the average reaction rate of the system.
This change reflects the effect of phase separation on the kinetics of chemical reactions.

\subsection{Bimolecular chemical reactions in coexisting phases}
\label{sec:second_order}

As a further example for a chemically reactive system at phase equilibrium,  we study a four component system which contains three reactive solutes $i=A,B,C$ and a non-reactive solvent $S$. In this example, the solutes undergo a bimolecular chemical reaction,
\be
\,A+ \,B  \rightleftharpoons \,C  \qc \label{eq:chem_scheme_bimolecule}
\ee
which conserves volume (see Appendix~\ref{app:example_quaternary} for details ).
For such a bimolecular chemical reaction in a four component mixture, there exists three conserved quantities. Each conserved quantity is represented by a plane in a three dimensional phase diagram spanned by the volume fractions of the reactive solute components $A,B,C$.
The three conserved quantities are $\psi_{0}=\nu_{S}n_{S}$, $\psi_{1}=\nu_{A}n_{A}+\nu_{B}n_{B}+\nu_{C}n_{C}$ and $\psi_{2}=\nu_{A}n_{A}-\nu_{B}n_{B}$ and the intersection of the planes corresponding to conserved quantities $\psi_{1}$ and $\psi_{2}$ yields a line in the phase diagram.

The chemical kinetics of a bimolecular reaction at phase equilibrium can be depicted as a chemical trajectory in the three dimensional phase diagram.
Figure~(\ref{fig:kinetics_quaternary}a) shows two chemical trajectories corresponding to systems with two different values of conserved quantities.
For both cases, the kinetics of the average composition follow the intersection of the respective conserved planes, $\psi_{1}$ and $\psi_{2}$, which is illustrated for one initial condition in Fig.~(\ref{fig:kinetics_quaternary}b).
As phase separation occurs the volume fractions in each phase move along the binodal surface (green).
The chemical kinetics stop when the volume fractions in the coexisting phases reach the thermodynamic equilibrium (green-orange dashed line in Fig.~(\ref{fig:kinetics_quaternary}a-b)).
The thermodynamic equilibria lie on a closed line is
given by the intersection between the binodal surface and the chemical equilibrium surface with the condition, $\mu_A+\mu_B=\mu_C$.
Concomitantly, the volume fraction of the product $C$ saturates (red and blue lines in Fig.~(\ref{fig:kinetics_quaternary}c)). These saturation values differ, since the thermodynamic equilibrium state depends on the conserved quantities (shown by the tie lines in Fig.~(\ref{fig:kinetics_quaternary}b)).

\begin{figure*}[t]
  \centering
    \includegraphics[width=0.99\textwidth]{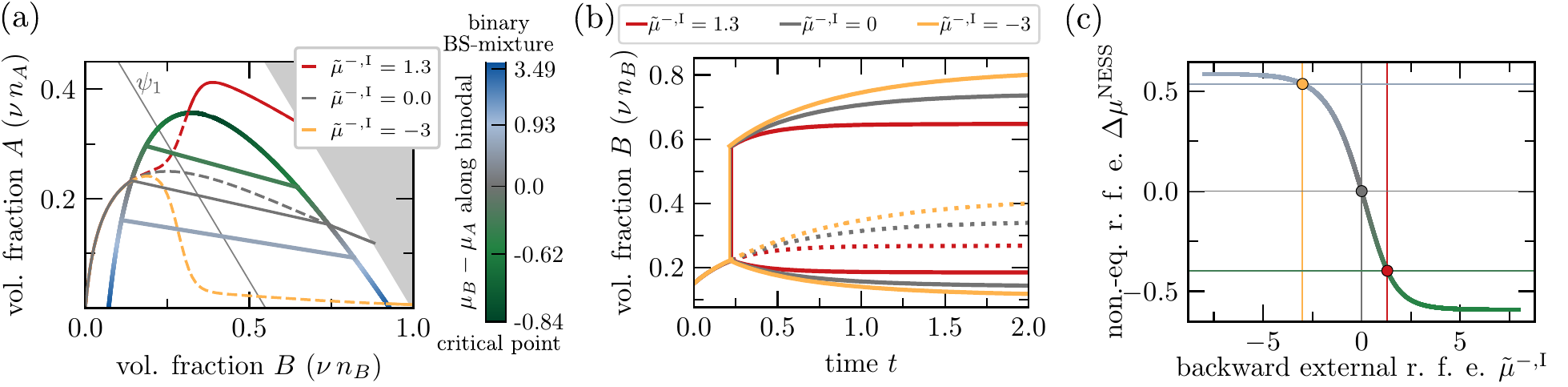}
    \caption{\textbf{Unimolecular chemical reaction maintained away from chemical equilibrium.}
    \textbf{(a)}
    Depending on the value of the backward external reaction free energy (r.f.e) $\tilde{\mu}^{-,\I}$, different coexisting phases are selected
    (green tie line for $\tilde{\mu}^{-,\I}=1.3$  and light blue gray line for $\tilde{\mu}^{-,\I}=-3.0$).
    The color bar represents the chemical potential difference between the components $B$ and $A$ along the binodal line, where the lower/upper bound corresponds to the critical point/binary $B$-$S$ mixture.
    \textbf{(b)} Chemical trajectories for the product \textit{B},
    where
    solid and dotted lines correspond to volume fractions in phase I/II and  averages, respectively. Chemical kinetics and non-equilibrium steady states vary with the value of the backward external r.f.e $\tilde{\mu}^{-,\I}$.
    \textbf{(c)}
    The more the backward external r.f.e $\tilde{\mu}^{-,\I}$ deviates from zero,
    the more the system is away from thermodynamic equilibrium, which is characterized by the non-equilibrium reaction r.f.e $\Delta\mu^\text{NESS}$.
    The gray dots correspond to the reference system at thermodynamic equilibrium, while the red and orange dots represent the two non-equilibrium steady states depicted in (a,b). Results in (a,b,c) are obtained for $\psi_{1}=0.55.$
    }
    \label{fig:active}
\end{figure*}

\subsection{Chemical reactions maintained away from chemical equilibrium}

Chemical reactions can also be maintained away from thermodynamic equilibrium. This is common in living cells, where biochemical reactions are chemically driven by the consumption of a chemical fuel.
Maintaining reactions away from chemical equilibrium can lead to non-equilibrium steady states (NESS) with non-vanishing diffusive exchange rates between the phases.

We consider systems at phase equilibrium where a fuel provides an external reaction free energy $\Delta \tilde{\mu}_{\alpha}=
\tilde{\mu}^-_\alpha - \tilde{\mu}^+_\alpha
$ that maintains the chemical reaction away from chemical equilibrium.
The reaction free energy then reads
\be \label{eq:mu_pm_NONTEQ}
 \Delta \mu_{\alpha} = \sum_{i=0}^{M}\sigma_{i\alpha} \mu_i +\Delta \tilde{\mu}_{\alpha} \qc
\ee
which can again be written as the difference of forward ($+$)  and backward ($-$)  reaction free energies
\be
\mu_\alpha^\pm = \sum_{i=0}^M \sigma_{i\alpha}^\pm \mu_i + \tilde{\mu}_\alpha^\pm \qd \label{eq:non_eq_mu_a_pm}
\ee
In certain cases, these systems reach an {effective} equilibrium state even if  $\Delta \tilde{\mu}_\alpha \neq 0$ and Eq.~\eqref{eq:mu_pm_NONTEQ} can be recasted in the form of Eq.~\eqref{eq:mu_pm_TEQ}: (i) $\Delta \tilde{\mu}_\alpha =\sum_{i=0}^M \sigma_{i\alpha} c_i$, where $c_i$ are constants and the system can phase-separate; (ii)  the chemical potential shifts $c_i(n_i)$ corresponding to $\Delta \tilde{\mu}_\alpha $ depending on composition but the system is spatially homogeneous.
Case (i) maintains thermodynamic consistency of phase and effective chemical equilibria, while case (ii) does not have phase equilibrium but  maintains effective chemical equilibrium~\cite{BEARD:2002}.
If these conditions are not satisfied, the system cannot reach a thermodynamic equilibrium.
Here, we focus on systems with composition dependent $\Delta \tilde{\mu}_{\alpha}$ in the presence of coexisting phases, thus neither case (i) nor case (ii) do apply.
Under these circumstances, the intersections between the binodal manifold with the effective chemical equilibrium manifold, $\Delta \tilde{\mu}_{\alpha}=0$ (e.g., yellow and red lines in Fig.~(\ref{fig:active}a)), will not be connected by a tie line. Therefore,
there can be non-equilibrium steady states, where $d n_i^\III/dt = 0$ and $d V^\III/dt=0$ with non-zero reaction rates $r_i^\III$ and diffusive exchange rates $j_i^\III$ between the coexisting phases.

A measure of the deviation of the chemical reaction $\alpha$ from thermodynamic equilibrium is the reaction free energy in the non-equilibrium steady state,
\be
\Delta \mu_\alpha^{\text{NESS}} \equiv  \sum_{i=0}^M \sigma_{i\alpha} \mu_{i}^{\text{NESS}} \qc
\label{eq:ness_def}
\ee
where $\mu_{i}^{\text{NESS}}$ are steady state chemical potentials. Since $\mu_i$ are identical in the two phases due to phase equilibrium, the non-equilibrium reaction free energy,
\begin{flalign}
&\begin{aligned}
\Delta &\mu_\alpha^{\text{NESS}}/k_BT = \\
&\log\left(\frac{k_\alpha^\I V^\I \exp\left({\frac{\tilde{\mu}_\alpha^{+,\I}}{k_BT}}\right)+k_\alpha^\II V^\II \exp\left({\frac{\tilde{\mu}_\alpha^{+,\II}}{k_BT}}\right)}{k_\alpha^\I V^\I \exp\left({\frac{\tilde{\mu}_\alpha^{-,\I}}{k_BT}}\right)+k_\alpha^\II V^\II \exp\left({\frac{\tilde{\mu}_\alpha^{-,\II}}{k_BT}}\right)} \right)
\end{aligned} \label{eq:ness}
\end{flalign}
is phase-independent. Equation~(\ref{eq:ness}) results from the balance of reaction and diffusive exchange rates $ r_i^\III = j_i^\III$, which corresponds to the steady state condition of \eq{dyn_dens} together with $dV^\III/dt=0$.
The non-equilibrium reaction free energies, $\Delta \mu_{\alpha}^{\text{NESS}}$ can be interpreted as susceptibilities of the system to external reaction free energies $\tilde{\mu}_\alpha^{\pm,\III}$. The values of the non-equilibrium reaction free energies are dependent on the reaction rate coefficients $k_\alpha^\III$ and the phase volumes at steady state, $V^\III$.

To illustrate the kinetics of chemical reactions and steady states that are maintained away from chemical equilibrium but are at phase equilibrium, we study the same unimolecular reaction in a ternary mixture as in Sec.~(\ref{sec:first_order}).
To maintain the reaction away from chemical equilibrium, we introduce a non-zero backward external reaction free energy $\tilde{\mu}^{-,\I}$.
For simplicity, the other reaction free energies $\tilde{\mu}^{\pm,\II}$ and $\tilde{\mu}^{+,\I}$ are chosen to be zero.
Therefore, the effective chemical equilibrium line, $\Delta \mu =0$, is only affected in phase I (solid red and yellow lines compared to the gray line in Fig.~(\ref{fig:active}a).
For a single reaction, however, a non-equilibrium steady state can only be reached in systems with coexisting phases.

For such systems, an important finding is
that by choosing different values of the external reaction free energy $
\tilde{\mu}^{-,\I}$, the chemical kinetics changes and the system relaxes to different non-equilibrium steady states.
For each value of $\tilde{\mu}^{-,\I}$,
such steady states have specific compositions in the coexisting phases
indicating that the chemical driving can select distinct states of the chemically reactive system.
(solid green and light blue, respectively, in Fig.~(\ref{fig:active}a-b)).
Moreover,
the chemical trajectories of the volume fractions in each phase (solid lines) and the average volume fractions (dotted lines) change when varying the external reaction free energy $\tilde{\mu}^{-,\I}$; see  Fig.~(\ref{fig:active}b).
In particular, the jump of the average acceleration $d\bar{r}_i/dt$ is also affected by the external reaction free energy (not shown).

The reaction free energy in the non-equilibrium steady state $\Delta\mu^{\text{NESS}}$ is used to characterise how much the considered system
deviates from thermodynamic equilibrium for a given value of the backward external reaction free energy $\tilde{\mu}^{-,\I}$.
Around thermodynamic equilibrium, $\Delta\mu^{\text{NESS}}(\tilde{\mu}^{-,\I})$ varies linearly, while for large deviations, $\Delta\mu^{\text{NESS}}$ saturates at two plateaus depending on the sign of the backward external reaction free energy $\tilde{\mu}^{-,\I}$ (Fig.~(\ref{fig:active}c)).
In particular, large and positive $\tilde{\mu}^{-,\I}$ favor the conversion from component $B$ to $A$ in phase I. This trend is opposed by a decrease in the volume of phase I, therefore leading to the plateau for $\tilde{\mu}^{-,\I}$.
Consistent with this, the value of the plateau of $\Delta\mu^{\text{NESS}}$ is determined by the volume of phase I which is in turn set by the conserved
quantity $\psi_{1}$.
Specifically, the plateau value corresponds to the intersection of the line of conserved quantity $\psi_{1}$ (thin gray line) and the binodal line; see Fig.~(\ref{fig:active}a).
In contrast, small and negative values of  $\tilde{\mu}^{-,\I}$
favor the  conversion from component $A$ to $B$
in phase I.
The full conversion is not possible
since chemical reactions are only maintained away from chemical equilibrium in phase I, and thus the volume of phase I limits
the selection of
of coexisting, non-equilibrium steady states.

For chemically driven systems,
the phase volumes are strongly influenced by an external supply of reaction free energy.
This property is distinct to  chemically driven systems since at thermodynamic equilibrium, the phase volumes
are solely determined by the conserved quantity.
In particular, a ternary mixture with one chemical reaction at  thermodynamic equilibrium becomes an effective binary mixture of two conserved quantities.
As a result, varying these conserved quantities solely
changes the phase volumes (Fig.~(\ref{fig:equil}c)).
In contrast, for a chemically driven system,
changing the conserved quantities also affect phase composition.

\section{Discussion \& Outlook}

In our work, we developed a theory of the chemical kinetics in phase-separated mixtures at phase equilibrium.
For simplicity, we considered homogeneous phases which applies to the case were chemical reactions are slow compared to the phase separation kinetics. This includes systems where chemical reactions are rate limiting, which is typical of biological enzymes ~\cite{Bar_Even:2011}.
This separation of time-scales implies that the size of each phase is smaller than the
reaction-diffusion length-scales which are set by the reaction rate coefficients and the diffusion coefficients.
If these conditions are satisfied, we can consider the case of chemical reactions
in coexisting phases that are each homogeneous and well-mixed.

For systems with chemical reactions that are slow compared to the phase separation kinetics, we showed that the condition of phase equilibrium governs chemical equilibrium.
Therefore, chemical equilibrium in coexisting phases differs from the chemical equilibrium where all components are well-mixed. Furthermore, the kinetics of reactions approaching chemical equilibrium also differs between phase-separated and the corresponding well-mixed system.
We show that in a phase-separated system, the relaxation kinetics can be represented by a chemical trajectory of  time-dependent concentrations that move along the binodal manifold in the phase diagram.
We also find that conservation laws play an important role in phase-separated systems. Quantities conserved by the reactions define manifolds in composition space to which average compositions are confined. Such conservation manifolds, together with the manifolds of chemical and phase equilibria, govern the combined kinetics of reactions and phase separation.


Phase separation organising chemical reactions was suggested as an important concept to understand cellular biochemistry~\cite{Alberti2017,Banani2017, Shin2017,weber2021}.
In particular, phase-separated condensates can provide distinct biochemical environments and serve to localise and confine chemical reactions~\cite{su:2016,SHEUGRUTTADAURIA:2018}.
The study of biochemical processes in phase-separated systems is currently a rapidly growing field.
Our theory can play an important role to interpret observations in experimental systems were solute components undergo chemical reactions in the presence of coexisting phases.
In particular, our work clarifies that the increased concentration of reactants in a condensed phase does not by itself lead to increased reaction rates.
Rather, if the coexisting phases are at phase equilibrium, the reaction rates $r_i^\III$ of component $i$ in each phase can only differ due to different reaction rate coefficients $k_{\alpha}^\III$.
In other words, the increased local concentration of a reactive solute due to phase separation does not necessarily increase the rates of reactions in which it participates.
The speed-up or slow-down of reactions is solely determined by the reaction rate coefficients in each phase, which can also decrease upon condensation.
These insights might be relevant to explain recent observations in coacervate emulsions with enzymatic  reactions~\cite{Drobot2018,Koga:2011,Kojima:2018,STROBERG:2018}.
Another important insight of our work is that the rate of change of the concentration of reactive molecule $i$ in one phase is not equal to the reaction rate $r_i^\III$ of this component.
This is because phases are coupled and components are rapidly exchanged between the phases at phase equilibrium.
To determine the reaction rate $r_i^\III$ of component $i$, the exchange rate between the phases as well as the changes in phase volumes need to be taken into account.
{Thus, the chemical kinetics in coexisting phases tightly integrates
phase separation kinetics and reaction kinetics.}
To highlight this point, we note that the effect of phase separation on chemical reactions in two coexisting phases cannot be inferred from the study of reactions in the two phases when they are isolated.

%

We also discussed chemical reactions at phase equilibrium but maintained away from chemical equilibrium via an external supply of free energy.
Such an external free energy could for instance be supplied via a chemical fuel.
We find that the resulting non-equilibrium steady states have non-zero reaction rates and diffusive transport of components between the phases.
We showed that the steady state concentrations in the two phases depend on the external reaction free energy.
Thus, controlling the external reaction free energy supply can be used to select distinct compositions of coexisting phases {and to vary the phase volumes.}
For systems maintained away from chemical equilibrium,
the reaction rates $r_i^\III$ can be phase-dependent due to the supply of external free energy which can differ between the phases, in addition to phase-dependent reaction rate coefficients $k^\III$.
Thus, for such driven systems, reaction rates in the two phases can be controlled externally,
even allowing opposite net directions of chemical reactions between the phases.

Using equations for the chemical kinetics of dilute, homogeneous mixtures for systems that can
phase-separate is incorrect if the system can relax toward thermodynamic equilibrium.
First, there are diffusive exchanges between the phases.
Second,
systems that phase-separate are non-dilute
and have activities that are non-linear in concentrations.
Such non-linear activities govern the kinetics of chemical reaction,
in particular at concentrations where mixtures can phase-separate.
For instance, in phase-separated systems, unimolecular reactions cannot be described as first order reactions when the system can relax toward thermodynamic equilibrium.
Applying
equations for the chemical kinetics of dilute, homogeneous mixtures to phase-separated systems would implicitly correspond to a driven system with a supply of external free energy.

The results of our work could be tested in experimental systems such as coacervates with enzymatic reactions~\cite{love:2020,Drobot2018,nakashima:2018,Altenburg:2020,chen:2020}.
It will be interesting to compare chemically reactive systems at phase equilibrium with their well-mixed counterparts.
An important extension of our work will be to account for the influence of interface between phases on chemical reactions.
In particular, reactive solute components at the interfaces could show a complex kinetics due interfacial effects
such as effective resistance~\cite{TAYLOR:2019,Hahn:2011,gebhard:2021}, thereby altering the transport between the coexisting phases.

\section*{Author Contributions}
All authors conceived the project
and contributed to the development of the theory.
J.B. and S.L. worked out the details of the theory,
and performed the numerical calculations.
All authors wrote the manuscript
and have given their approval to the final version of the manuscript.

\section*{Conflicts of Interest}
There are no conflicts to declare.

\acknowledgments{We thank Stefano Bo,
Giacomo Bartolucci and Tyler Harmon for insightful discussions.
We kindly thank Evan Spruijt for providing helpful references on the history of the law of mass action.
F.\ J\"ulicher
acknowledges funding by the Volkswagen Foundation.
C.\ Weber acknowledges the European Research Council (ERC) under the European Union’s Horizon 2020 research and innovation programme (``Fuelled Life'' with Grant agreement No.\ 949021) for financial support.
}

\appendix

\section{Gibbs free energy density for multicomponent mixture}
\label{app:Gibbs_energy}
The Gibbs free energy density for $M+1$ components is as follows,
\begin{equation}
\label{eq:free_energy}
\begin{split}
    g(\{n_{i}\},p,T)&=k_{B}T\sum^{M}_{i=0}n_{i} \log(\nu_{i}n_{i})+\sum^{M}_{\substack{i,j=0}}\frac{\chi_{ij}}{2}n_{i}n_{j}\\
    &+\sum^{M}_{i=0}\omega_{i}(T)n_{i} + p.
    \end{split}
\end{equation}
The chemical potential of each component is therefore as previously defined,
\be
\begin{split}
\mu_i &\equiv \partial G/ \partial N_i|_{T,p,N_{i\neq j}} \\&= \frac{\partial g}{\partial n_i}+\nu_i \bigg(g-\sum^{M}_{k=0} \frac{\partial g} {\partial n_k} n_k\bigg)
\end{split}
\ee

\be
\begin{split}
\mu_i=&\nu_{i}p+\omega_{i}+k_{B}T(1+\ln(n_{i}\nu_{i}))+\sum^{M}_{k=0}\chi_{ik}n_{k}\\&-\nu_{i}\bigg(\sum^{M}_{\substack{k,l=0\\k\neq l}}\frac{\chi_{kl}}{2}n_{k}n_{l}+k_{B}T\sum^{M}_{k=0} n_{k}\bigg)
\end{split}
\ee
Using this expression of chemical potential, we can re-write it as in Eq.~\eqref{eq:chem_pot} and therefore obtain the activity coefficients as in Eq.\eqref{eq:gamma_specific} in the main text.

\section{Kinetics of particle number and phase volumes}\label{app:volume_kinetics}
The rate of change of particle number $N_i$ in each phase due to chemical reactions occurring with the rate $R_{i}$ and the exchange rate $J_{i}$ between the phases $J_{i}$ is given by,
\be
\label{eq:dyn_particlenumber}
\frac{d}{dt} N_i^\III = R_i^\III - J_i^\III.
\ee
Due to conservation of number of particles of individual components during the exchange between phases, $J_i^I=-J_i^\II$ and
\be
\label{eq:reactionrate}
R_i^\III = V^\III r^\III_{i} \qc
\ee
where
$r^\III_{i}= \sum^{R}_{\alpha=1} \sigma_{i\alpha}  r_\alpha^\III$ and
$r_\alpha=(r_\alpha^+ - r_\alpha^-)$, each with the units of particle number per time and volume.
Each phase volume is defined as the total volume occupied by all the components in each phase, $V^\III=\sum^{M}_{i=0} \nu_i N_i^\III$. Therefore, the dynamic equation for the phase volume is,
\be
 \frac{d}{dt} V^\III
= \sum^{M}_{i=0} \nu_i^\III ( R_i^\III - J_i^\III) + \sum_i N_i^\III\frac{d}{dt}\nu_i^\III \qd
\ee
Due to incompressibility, $d\nu^\III_{i}/dt=0$ and the molecular volumes are constant and identical in each phase. Dividing both sides of the  equation above with the respective phase volumes and using Eq.~\eqref{eq:reactionrate}, we obtain Eq.~\eqref{eq:dyn_phasevol}.
We can now derive the dynamic equations Eq.~\eqref{eq:dyn_dens} for the concentrations $n^{\III}_{i}$ using Eq.~\eqref{eq:dyn_particlenumber} and Eq.~\eqref{eq:reactionrate}.

\section{Parameters used for numerical computations}
\label{app:example_ternary}

\subsection{Unimolecular chemical reaction}
Given three components and one chemical reaction, we define two conserved quantities, $\psi_{0}=\nu_{S} n_{S}$ and $\psi_1=(\nu_{A}n_A+\nu_{B}n_{B})$.
The kinetic equations for the reactive solutes in each phase are obtained using Eq.~\eqref{eq:dyn_dens}. We choose the initial average volume fractions along conserved  line $\psi_{1}=0.55$ for kinetic studies in Fig.~(\ref{fig:kinetics}).

The thermodynamic parameters used to construct Fig.~(\ref{fig:equil}) and Fig.~(\ref{fig:kinetics}) are listed below.
The reference chemical potentials in units of $k_{B}T$ are $\mu^{0}_{A}/(k_{B}T)=2$ and $\mu^{0}_{B}/(k_{B}T)=0$. The molecular volumes of all components, in units of solvent molecular volume $\nu_{S}$ are identical $\nu_A/\nu_{S}=1$ and $\nu_B/\nu_{S}=1$. Therefore we use $\nu$ for the identical molecular volumes.
The mean field interaction parameters in units of $(k_{B}T \nu_{S})$ are
$\chi_{AS}/(k_BT\nu_S)=-1$, $\chi_{BS}/(k_BT\nu_S)=3$ and $\chi_{AB}/(k_BT\nu_S) = 0$.
The reaction rate coefficient used for studying the reaction kinetics of the system in Fig.~(\ref{fig:kinetics}a-b) is chosen identical in each phase, $k^\III=1$ per unit time.
In Fig.~(\ref{fig:kinetics}c-f), for different ratios of reaction rate coefficients, the value of $k^\II=1$ per unit time is kept constant and $k^\I$ is varied accordingly to maintain the ratio.
The initial conditions for two cases in Fig.~(\ref{fig:kinetics}a-b) are (i) $\nu \bar{n}_{A}(0)=0.4$, $\nu \bar{n}_{B}(0)=0.15$ and (ii) $\nu \bar{n}_{A}(0)=0.05$, $\nu \bar{n}_{B}(0)=0.5$.

\subsection{Bimolecular chemical reaction}
\label{app:example_quaternary}

Given four components and one chemical reaction, we have three conserved quantities as defined in Sec.~(\ref{sec:second_order}).
The kinetic equations for the reactive solutes in each phase are obtained using Eq.~\eqref{eq:dyn_dens}.
We choose the initial average volume fractions corresponding to two systems along intersection of conserved plane $\psi_{1}=0.42$ and $\psi_{2}=0.05$ and $0.1$, respectively.
The thermodynamic parameters used to construct Fig.~(\ref{fig:kinetics_quaternary}) are listed below. The reference chemical potentials in units of  $k_{B}T$ are  $\mu^{0}_{A}/(k_{B}T)=3.5$, $\mu^{0}_{B}/(k_{B}T)=3.5$ and $\mu^{0}_{C}/(k_{B}T)=0$. The molecular volumes in units of solvent volume fraction $\nu_S$ are $\nu_{A}/\nu_S=1$, $\nu_{B}/\nu_{S}=1.0$ and $\nu_C/\nu_{S}=2$, respectively. The mean field interaction parameters in units of $(k_{B}T \nu_{S})$ are  $\chi_{AS}/(k_{B}T \nu_{S})=1.25$, $\chi_{BS}/(k_{B}T \nu_{S})=0$,  $\chi_{CS}/(k_{B}T \nu_{S})=3.2$, $\chi_{AB}/(k_{B}T \nu_{S}) = 0$, $\chi_{AC}/(k_{B}T \nu_{S}) = -1.25$ and $\chi_{BC}/(k_{B}T \nu_{S}) = 0$.

The reaction rate coefficient is chosen identical in each phase, $k^\III=1$ per unit time.

\subsection{Unimolecular chemical reaction maintained away from chemical equilibrium}
\label{app:example_ternary_active}
The thermodynamic parameters used to construct Fig.~(\ref{fig:active}) are the same as in Appendix~(\ref{app:example_ternary}1). We choose the conserved quantity as $\psi_{1}=0.55$ (Figs.~(\ref{fig:active}a-c)). The initial condition used in Fig.~(\ref{fig:active}b) is $\nu \bar{n}_{A}(0)=0.4$ and $\nu \bar{n}_{B}(0)=0.15$, which lies in the mixed phase. We choose the value of backward external reaction free energy in the mixed phase from the sigmoidal function which depends on the product $B$ volume fraction,
\be
\tilde{\mu}^{-}\left( n_{B}\right)=0.5\, \bigg[\text{tanh}\bigg(\frac{\nu \, n_{B}-0.3}{0.05}\bigg)+1 \bigg]\tilde{\mu}^{-,\I} \qd
\ee,
where $\tilde{\mu}^{-,\I}$ is fixed to two different non-zero values (see legends in Fig.~(\ref{fig:active})).
The choice of the external reaction free energies alongwith the conserved quantity uniquely defines the NESS composition in the coexisting phases.

\newpage
\bibliography{lib}
\end{document}